\documentclass{aastex}
\usepackage{spr-astr-addons}
\usepackage{url}
\usepackage{graphicx,epsfig}
\usepackage{amsmath}
\RequirePackage{color}

\def\spose#1{\hbox to 0pt{#1\hss}}
\def\lta{\mathrel{\spose{\lower 3pt\hbox{$\mathchar"218$}}
     \raise 2.0pt\hbox{$\mathchar"13C$}}}
\def\gta{\mathrel{\spose{\lower 3pt\hbox{$\mathchar"218$}}
     \raise 2.0pt\hbox{$\mathchar"13E$}}}

\begin{document}

\setlength{\overfullrule}{0pt}

\title{
On the age and metallicity of planet-hosting triple star systems
}

\shorttitle{Age and metallicity of triple star systems}
\shortauthors{Cuntz \& Patel}

\author{M. Cuntz\altaffilmark{1}} \and \author{S. D. Patel\altaffilmark{1}}


\altaffiltext{1}{Department of Physics, University of Texas at Arlington, Box 19059,
       Arlington, TX 76019, USA. \\
       email: cuntz@uta.edu \\
       email: shaan.patel@uta.edu
}

\begin{abstract}
  We present a statistical analysis of the ages and metallicities of
  triple stellar systems that are known to host exoplanets.  With
  controversial cases disregarded, so far 27 of those systems have been
  identified.  Our analysis, based on an exploratory approach,
  shows that those systems are on average
  notably younger than stars situated in the solar neighborhood.
  Though the statistical significance of this result is not fully established,
  the most plausible explanation is a possible double selection effect
  due to the relatively high mass of planet-hosting stars of those
  systems (which spend less time on the main-sequence than low-mass
  stars) and that planets in triple stellar systems may be long-term
  orbitally unstable.  The stellar metallicities are on average
  solar-like; however, owing to the limited number of
  data, this result is not inconsistent with the previous finding that
  stars with planets tend to be metal-rich as the deduced
  metallicity distribution is relatively broad.
\end{abstract}

\keywords{astrobiology -- binaries: general -- methods: statistical -- planetary systems --
stars: fundamental parameters -- stars: late-type}


\section{Introduction}

For more than a decade, the concept of habitable planets in binary and multiple stellar
systems has been a topic of great interest, fueled by both significant observational
and theoretical advances.  Previous observations indicate that a large percentage 
of stellar binary and higher order systems exist in our galactic neighborhood, including
systems harboring planets \citep[e.g,][]{duqu91,pati02,egge04,ragh06,ragh10,roel12}.
Those systems provide a unique set of candidates in the search for life beyond Earth.  

Conforming cases include planet-hosting triple star systems, which are both profoundly
interesting and highly challenging from both observational and theoretical perspectives,
especially regarding aspects of planet formation \citep[e.g.,][]{domi15} and planetary
orbital stability \citep[e.g.,][]{geor13,corr16,buse18,myll18,boyl21}.  A catalog of
planet-harboring triple stellar systems has been published by \cite{cunt22}.  They
pointed out that planets in triple star systems exhibit a large variety, including
sub-Earths, Earth-type planets, super-Earths, as well as Neptune and Jupiter-type
planets; the latter are by far in the majority.  Most of these objects have been
discovered via the Radial Velocity or the Transit method.

In this work, we pursue a follow-up study, aimed at examining the age and
metallicity distribution of stars that are members of planet-hosting triple star systems;
this work is based on results available in the literature.  For the determination
of stellar ages, various methods have been adopted, including isochrone fitting,
chromospheric and X-ray emission, and rotation and lithium depletion analyses, as,
e.g., for HD~132563; see \cite{desi11} and references therein.

Stellar metallicity determinations are based on high-quality spectroscopy.  This
also allows comparisons with previous work for single stars, which on average
are found to be metal-rich \citep[e.g.,][~and related work]{gonz97,gonz98,fisc05}
Although the number of systems with reliable age and metallicity determinations
is relatively small (26 each), a tentative statistical analysis for the current sample
of stars can be pursued, which however should be updated in the future when
further observations have become available.

Our paper is structured as follows: Section~2 conveys examples of planet observations,
especially for the listing of systems considered here.  Comments on the acquisition of
data are given in Section~3.  In Section~4, we report on the statistical analysis for
the age and metallicity distribution of planet-hosting triple star systems, the focus
of this study.  Section~5 and 6 convey our discussion, and the conclusions and outlook,
respectively.


\section{Exoplanet Observations}

To date, 27 planet-hosting triple star systems have been identified not including
controversial or retracted cases (see Table~1).  All those systems are highly hierarchic,
consisting of a binary complemented by a distant stellar component in orbit about
the common center of mass; see \cite{buse18} and \cite{cunt22} for a summary and
evaluation of previous results.

Regarding the binary, both S-type and P-type orbits are possible; see \cite{dvor82} as
source of nomenclature.  In case of S-type orbits, planets are in orbit
about one of the binary components, whereas in case of P-type orbits, planets are
found in orbit about both components; see \cite{cunt14} for a comprehensive approach.
If the planet is hosted by the distant stellar component, another kind of S-type orbit
is established.

The best studied case of planets hosted by triple star systems is that of
Alpha Centauri.  This system is closest to Earth, and any exoplanets hosted
by that system are closest to Earth as well, regardless of any future planet discoveries.
So far, three planets have been identified (with one planet still mildly controversial).
They are all hosted by the outlying component also known as Proxima Centauri;
see \cite{angl16}, \cite{dama20}, and \cite{fari22} for further information.

Arguably, the first planet discovered harbored by a triple star system is 16~Cyg~Bb.
It is a Jupiter-mass planet found by \cite{coch97}.  However, at that time the outlying
stellar component, i.e., 16~Cyg~C, a red dwarf, was still unknown; evidence on
its existence has later been published by \cite{haus99}.  This subsequent result
elevated 16~Cyg to a triple stellar system with a planet.

Most planets in planet-hosting triple star systems have been discovered based on the
Radial Velocity (RV) method, with the Transit method, largely associated with the
{\it Kepler} mission, as a distant second.  Key facilities of those planetary
RV discoveries include the Lick, Keck, La Silla, and McDonald observatories, which
allowed for high-precision spectroscopy; see \cite{cunt22} for detailed references
and additional comments.

Most planets currently known to exist in triple star systems are Jupiter-mass
planets, a well-understood selection effect.  However, Neptune and
Earth-mass planets as well as sub-Earths and super-Earths have been
found as well.  An intriguing example is Gliese 667C known to harbor
two super-Earths; see \cite{angl12}.  Recently, \cite{sloa23} argued in favor
of possible exolife in that system.  Hence, the possible relevance of planets
situated in multiple stellar systems to astrobiology is another motivation
to explore aspects of metallicity and stellar age, as done here.

%
\begin{table*}
	\centering
	\caption{Planet detections}
	\begin{tabular}{clcl}
\noalign{\smallskip}
\hline
\noalign{\smallskip}
Index & Name                  &  Planets  & Reference                    \\
\noalign{\smallskip}
\hline
\noalign{\smallskip}
1     & 16~Cygni              &  1        &  \cite{coch97}               \\
2     & 51~Eridani            &  1        &  \cite{maci15}               \\
3     & 91~Aquarii            &  1        &  \cite{mitc03}               \\
4     & 94~Ceti               &  1        &  \cite{mayo04}               \\
5     & Alpha~Centauri        &  3~(2)    &  \cite{angl16,dama20,fari22} \\
6     & Epsilon~Indi          &  1        &  \cite{feng19}               \\
7     & Gliese~667            &  2~(1)    &  \cite{angl12}               \\
8     & HAT-P-8               &  1        &  \cite{lath09}               \\
9     & HAT-P-57              &  1        &  \cite{hart15}               \\
10    & HD~126614             &  1        &  \cite{howa10}               \\
11    & HD~132563             &  1        &  \cite{desi11}               \\
12    & HD~178911             &  1        &  \cite{zuck02}               \\
13    & HD~185269             &  1        &  \cite{mout06}               \\
14    & HD~196050             &  1        &  \cite{jone02}               \\
15    & HD~2638 / HD~2567     &  1        &  \cite{mout05}               \\
16    & HD~40979              &  1        &  \cite{fisc03}               \\
17    & HD~41004              &  1        &  \cite{zuck04}               \\
18    & HD~65216              &  1        &  \cite{mayo04}               \\
19    & K2-290                &  2        &  \cite{hjor19}               \\
20    & Kelt-4                &  1        &  \cite{east16}               \\
21    & Kepler-13             &  1        &  \cite{shpo11}               \\
22    & Kepler-444            &  5        &  \cite{camp15}               \\
23    & KOI-5 = TOI-1241      &  1        &  \cite{hirs17}               \\
24    & LTT~1445              &  1        &  \cite{wint19}               \\
25    & Psi$^1$~Draconis      &  1        &  \cite{endl16}               \\
26    & WASP-8                &  2        &  \cite{quel10,knut14}        \\
27    & WASP-12               &  1        &  \cite{hebb09}               \\
\noalign{\smallskip}
\hline
\noalign{\medskip}
\multicolumn{4}{p{1.8\columnwidth}}{Note: See \cite{cunt22} for additional information on
the systems.
}
\end{tabular}
\end{table*}


\section{Data Acquisition}

\subsection{General Comments}

Regarding the data acquisition and the associated statistics, the work
follows a two-tier approach.  For the stellar ages and metallicities,
information is obtained from the literature.  Hence, the authors are
not engaged in any of those determinations but report existing information,
if available.  The two times 26 data, in reference to both the stellar
ages and metallicities, have been subsequently subjected to a statistical
analysis and discussion (see Section~4), the main focus of this work.

\subsection{Stellar Ages}

Table 2 and 3 list information on the ages of the stellar system 
components.  If information is available for multiple stellar components
(which is typically not the case), average values have been used for the
subsequent analysis.  The results for the various systems are depicted
in Fig. 1.  Figure 2 provides a histogram indicating the shape of the
overall distribution.  Data are available for all systems\footnote{
Both for the stellar ages and metallicities,
data are available for 26 out of 27 objects (while considering combined
data due to stellar multiplicity).  This insufficiency
is not expected to significantly affect the conclusions of this work.},
except for LTT~1445, resulting in data for 26 systems.

For stellar age determinations, a variety of methods has previously been
used, including isochrone fitting, chromospheric and X-ray relationships,
and lithium depletion analyses; the latter is applicable to young,
low-mass stars.  For example, \cite{desi04} utilized all of these methods
for HD~132563.  Previously, \cite{saff05} applied the chromospheric emission
relationship regarding Ca~II~H+K for the age determination of a large number
of exoplanet host stars, including HD~41004 considered here.  Other studies
targeting chromospheric emission --- stellar age relationships have been
given by, e.g., \cite{noye84}, \cite{bali95}, \cite{cunt99}, and \cite{engl24}.

Regarding the pivotal planet-hosting Alpha Centauri system, a detailed
age determination has been given by \cite{joyc18}, which is based on
classically and asteroseismically constrained stellar evolution models,
including detailed fits with observations.  They arrived at $5.3 \pm 0.3$~Gyr,
a value that is also consistent with other models, including those by
\cite{kerv17} and \cite{vian18}.

Note that the youngest system considered in our study is 51~Eri, for which
26 $\pm$ 3 Myr is used \citep{niel16}.  A very similar value, which
is 24 $\pm$ 3 Myr, has been given by \cite{bell15}.  As pointed out by those
authors, 51~Eri belongs to a nearby moving group in the solar neighborhood,
thus allowing reliable absolute isochronal age determinations.

%
\begin{table*}
	\centering
	\caption{Age of stellar systems}
	\begin{tabular}{clcl}
\noalign{\smallskip}
\hline
\noalign{\smallskip}
Index & Name         &  Age    & Reference \\
...   & ...          &  (Gyr)  & ...       \\
\noalign{\smallskip}
\hline
\noalign{\smallskip}
1       & 16~Cygni              & 6.91 $\pm$ 0.26                       & {\tt see Table~3} \\
2       & 51~Eridani            & 0.026 $\pm$ 0.003                     & \cite{niel16}  \\
3$^a$   & 91~Aquarii            & 2.98 $\pm$ 1.63                       & \cite{bain18}  \\
4$^b$   & 94~Ceti               & $\sim$4.8                             & \cite{boya13}  \\
5       & Alpha~Centauri        & 5.261 $\pm$ 0.946                     & \cite{joyc18}  \\
6       & Epsilon~Indi          & 4.7 $\pm$ 1.0                         & \cite{feng19}  \\
7$^a$   & Gliese~667            & 6 $\pm$ 4                             & {\tt see Table~3} \\
8       & HAT-P-8               & 4.3$\substack{+1.5 \\ -1.4}$          & \cite{manc13}  \\
9$^a$   & HAT-P-57              & 1.0$\substack{+0.67 \\ -0.51}$        & \cite{hart15}  \\
10      & HD~126614             & 7.2 $\pm$ 2.0                         & \cite{vale05}  \\
11$^b$  & HD~132563             & $\sim$5                               & \cite{desi11}  \\
12      & HD~178911             & 4.8 $\pm$ 1.3                         & \cite{bonf16}  \\
13      & HD~185269             & 3.40 $\pm$ 0.54                       & \cite{jofr15}  \\
14$^a$  & HD~196050             & 2.5 $\pm$ 1.3                         & \cite{chav19}  \\
15$^a$  & HD~2638 / HD~2567     & 1.9 $\pm$ 2.6                         & \cite{bonf15}  \\
16$^a$  & HD~40979              & 1.92 $\pm$ 1.08                       & \cite{jofr15}  \\
17$^a$  & HD~41004              & 1.56 $\pm$ 0.8                        & \cite{saff05}  \\
18      & HD~65216              & 1.7 $\pm$ 0.5                         & \cite{bonf15}  \\
19      & K2-290                & 4.0$\substack{+1.6 \\ -0.8}$          & \cite{hjor19}  \\
20      & Kelt-4~A              & 4.44$\substack{+0.78 \\ -0.89}$       & \cite{east16}  \\
21$^a$  & Kepler-13             & 0.5 $\pm$ 0.1                         & \cite{shpo14}  \\
22      & Kepler-444            & 11.23$\substack{+0.91 \\ -0.99}$      & \cite{camp15}  \\
23      & KOI-5 = TOI-1241      & 3.49 $\pm$ 0.41                       & \cite{bell19}  \\
24      & LTT~1445              &    ...                                & ...            \\
25      & Psi$^1$~Draconis      & 2.4 $\pm$ 0.3                         & {\tt see Table~3} \\
26      & WASP-8                & 0.3$\substack{+0.9 \\ -0.0}$          & \cite{sout20}  \\
27$^a$  & WASP-12               & 2.0 $\pm$ 1.0                         & \cite{hebb09}  \\
\noalign{\smallskip}
\hline
\noalign{\medskip}
\multicolumn{4}{p{1.3\columnwidth}}{Note: Error bars assumed as 1$\sigma$ except if noted otherwise.
$^a$Error bar assumed as 3$\sigma$.
$^b$Error bar estimated as $\pm$1.0.
}
\end{tabular}
\end{table*}

%
\begin{table*}
	\centering
	\caption{Age of stellar components}
	\begin{tabular}{clcl}
\noalign{\smallskip}
\hline
\noalign{\smallskip}
Index & Name         &  Age    & Reference \\
...   & ...          &  (Gyr)  & ...       \\
\noalign{\smallskip}
\hline
\noalign{\smallskip}
1     & 16~Cygni~A            & 7.07 $\pm$ 0.26               & \cite{metc15}  \\
1     & 16~Cygni~B            & 6.74 $\pm$ 0.24               & \cite{metc15}  \\
7     & Gliese~667~AB         & 2--10                         & \cite{cayr81}  \\
7     & Gliese~667~C          & $>$2                          & \cite{cayr81}  \\
25    & Psi$^1$~Draconis~A    & 2.3 $\pm$ 0.3                 & \cite{endl16}  \\
25    & Psi$^1$~Draconis~B    & 2.5 $\pm$ 0.3                 & \cite{endl16}  \\
\noalign{\smallskip}
\hline
\noalign{\medskip}
\multicolumn{4}{p{1.3\columnwidth}}{Note: Error bars assumed as 1$\sigma$.
}
\end{tabular}
\end{table*}

\begin{figure}[ht]
    \centering
	\includegraphics[width=\columnwidth]{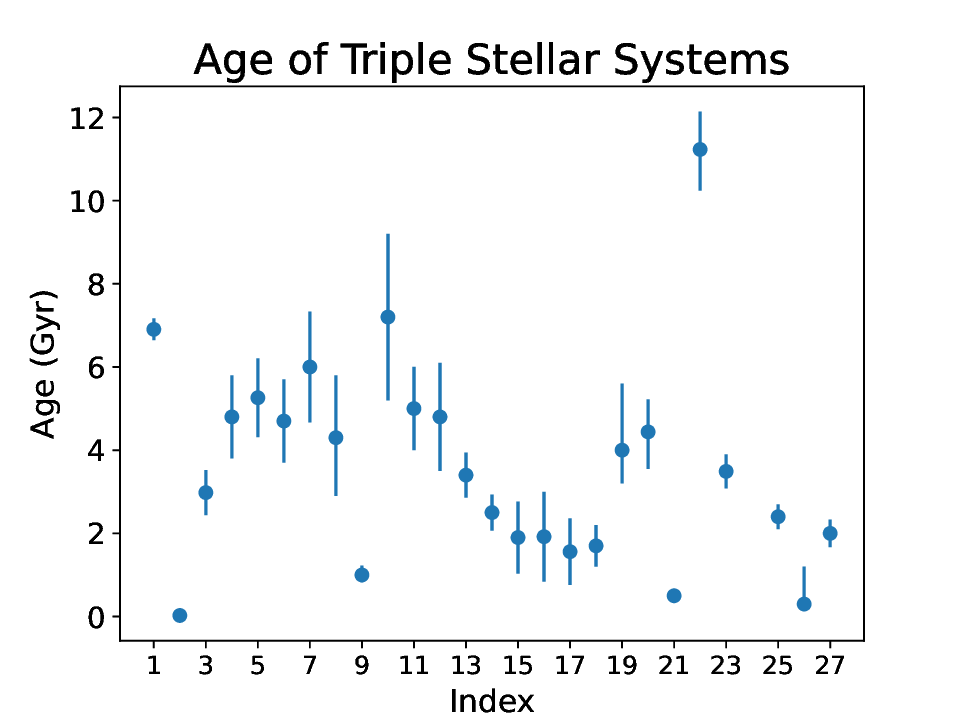}
    \caption{Ages of triple stellar systems, with average values
for the stellar components used if multiple determinations are available.
All error bars are 1$\sigma$.
}
\end{figure}

\begin{figure}[ht]
    \centering
	\includegraphics[width=\columnwidth]{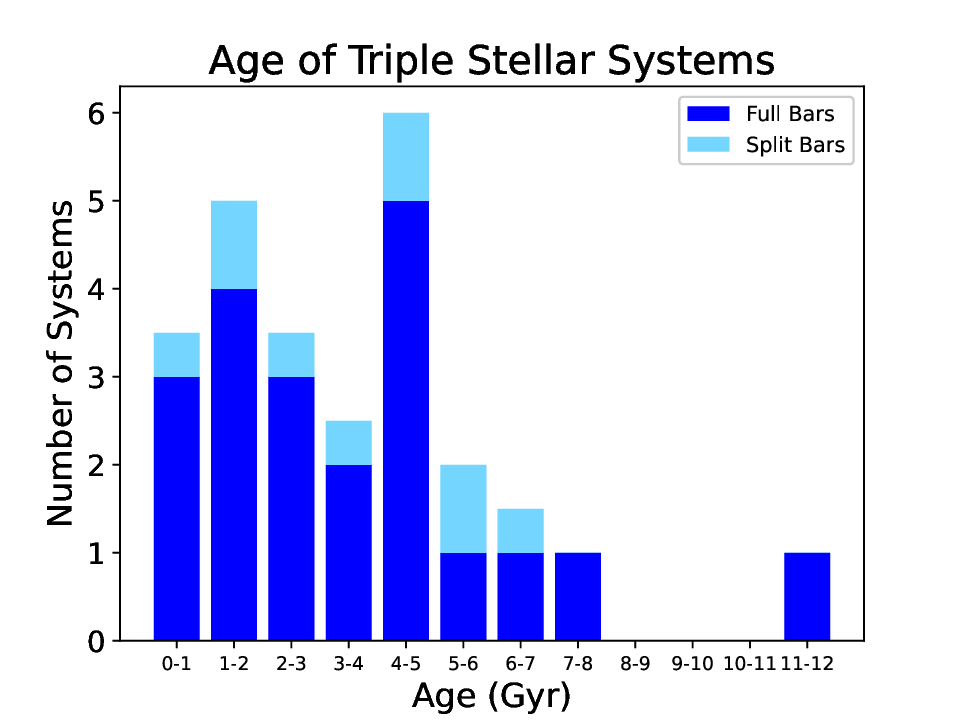}
    \caption{Histogram for the age distribution of
      planet-hosting triple star systems based on the
      average values for the stellar components.  Split
      bars are used to depict border values between bins.
}
\end{figure}

\subsection{Stellar Metallicities}

Table 4 and 5 list information on the metallicity of the stellar system 
components.  If information is available for multiple stellar components
(which, again, is typically not the case), average values have again been
used for the subsequent analysis.  The results for the various systems
are depicted in Fig. 3.  Figure 4 provides a histogram indicating the
shape of the overall distribution.  Data are available for all systems,
except for KOI-5, resulting in data for 26 systems.

Metallicity determinations ultimately require quality spectroscopy.
In some cases, those values are given as part of the associated planet
discovery publication.  Examples include Gliese 667 \citep{angl12},
HD~185269 \citep{mout06}, Kepler-13 \citep{shpo11}, and
Kepler-444 \citep{camp15}.

For the Alpha Centauri system, information on the stellar metallicities
have been given by \cite{schl10} and \cite{more18}; the results for the
three stellar components agree well within the uncertainty bars.  The
latter authors studied Alpha Cen~AB (whose masses bracket that of the
Sun) regarding a large number of elements.  Furthermore, \cite{schl10}
studied Alpha Cen~C, the planet-hosting component of the system as part
of a larger sample of M dwarfs.  They found that its metallicity
consistent with those of Alpha Cen~AB.

16~Cyg~A and B, with 16~Cyg~B previously identified as the hub of the
planet, see \cite{coch97}, has been studied by \cite{tucc14}.  They
concluded that 16~Cyg~A has a metallicity ([Fe/H]) higher by 0.047
$\pm$ 0.005 dex than 16~Cyg~B; this kind of difference has been
identified for all elements under consideration.  This result has been
interpreted as a possible signature of the rocky accretion core
of the giant planet 16~Cyg~Bb.

Another example is the work by \cite{desi04} who studied abundance
differences between components of wide binaries, including HD~132563,
based on a line-by-line differential analysis.  HD~132563Bb, a
Jupiter-type planet in an orbit of moderate eccentricity, was later
discovered by a team led by the same author.

%
\begin{table*}
	\centering
	\caption{Metallicity of stellar systems}
	\begin{tabular}{clcl}
\noalign{\smallskip}
\hline
\noalign{\smallskip}
Index & Name         &  Metallicity & Reference \\
...   & ...          &  (dex)       & ...       \\
\noalign{\smallskip}
\hline
\noalign{\smallskip}
1      & 16~Cygni              & $+$0.078 $\pm$ 0.025                  & {\tt see Table~5} \\
2      & 51~Eridani            & $-$0.12  $\pm$ 0.06                   & \cite{raja17} \\
3$^a$  & 91~Aquarii            & $-$0.14  $\pm$ 0.05                   & \cite{mass08} \\
4      & 94~Ceti               & $+$0.15  $\pm$ 0.07                   & \cite{fuhr08} \\
5      & Alpha~Centauri        & $+$0.23  $\pm$ 0.03                   & {\tt see Table~5} \\
6      & Epsilon~Indi          & $-$0.13  $\pm$ 0.06                   & \cite{delg17} \\
7      & Gliese~667            & $-$0.59  $\pm$ 0.10                   & \cite{angl12} \\
8      & HAT-P-8               & $-$$0.018\substack{+0.072 \\ -0.056}$ & \cite{wang21} \\
9$^b$  & HAT-P-57              & $-$0.25 $\pm$  0.25                   & \cite{hart15} \\
10     & HD~126614             & $+$0.56 $\pm$  0.04                   & \cite{vale05} \\
11     & HD~132563             & $-$0.185 $\pm$ 0.10                   & {\tt see Table~5} \\
12$^a$ & HD~178911             & $+$0.23 $\pm$  0.05                   & \cite{luck17} \\
13     & HD~185269             & $+$0.10 $\pm$  0.08                   & \cite{mout06} \\
14     & HD~196050             & $+$0.34 $\pm$  0.06                   & \cite{chav19} \\
15     & HD~2638 / HD~2567     & $+$0.12 $\pm$  0.05                   & \cite{tsan13} \\
16     & HD~40979              & $+$0.20 $\pm$  0.03                   & \cite{luck17} \\
17     & HD~41004              & $+$0.15 $\pm$  0.03                   & \cite{sous18} \\
18$^a$ & HD~65216              & $-$0.17 $\pm$  0.05                   & \cite{adib12} \\
19     & K2-290                & $-$0.06 $\pm$  0.10                   & \cite{hjor19} \\
20     & Kelt-4                & $-$$0.116\substack{+0.065 \\ -0.069}$ & \cite{east16} \\
21$^b$ & Kepler-13             & $+$0.2  $\pm$  0.2                    & \cite{shpo14} \\
22     & Kepler-444            & $-$0.69 $\pm$  0.09                   & \cite{camp15} \\
23     & KOI-5 = TOI-1241      &         ...                           & ...           \\
24     & LTT~1445              & $-$0.34 $\pm$  0.08                   & \cite{wint19} \\
25     & Psi$^1$~Draconis      & $-$$0.05\substack{+0.05 \\ -0.07}$    & {\tt see Table~5} \\
26     & WASP-8                & $+$0.29 $\pm$  0.03                   & \cite{mort13} \\
27     & WASP-12               & $+$0.21 $\pm$  0.04                   & \cite{mort13} \\
\noalign{\smallskip}
\hline
\noalign{\medskip}
\multicolumn{4}{p{1.3\columnwidth}}{Note: Error bars assumed as 1$\sigma$ except if noted otherwise.
$^a$Error bar estimated as $\pm$0.05.
$^b$Error bar assumed as 3$\sigma$.
}
\end{tabular}
\end{table*}

%
\begin{table*}
	\centering
	\caption{Metallicity of stellar components}
	\begin{tabular}{clcl}
\noalign{\smallskip}
\hline
\noalign{\smallskip}
Index & Name         &  Metallicity & Reference \\
...   & ...          &  (dex)       & ...       \\
\noalign{\smallskip}
\hline
\noalign{\smallskip}
1     & 16~Cygni~A            & $+$0.101 $\pm$ 0.008 & \cite{tucc14}  \\
1     & 16~Cygni~B            & $+$0.054 $\pm$ 0.008 & \cite{tucc14}  \\
5     & Alpha~Centauri~A      & $+$0.22  $\pm$ 0.02  & \cite{more18}  \\
5     & Alpha~Centauri~B      & $+$0.24  $\pm$ 0.03  & \cite{more18}  \\
5$^a$ & Alpha~Centauri~C      & $+$0.21  $\pm$ 0.05  & \cite{schl10}  \\
11    & HD~132563~A           & $-$0.18  $\pm$ 0.10  & \cite{desi04}  \\
11    & HD~132563~B           & $-$0.19  $\pm$ 0.10  & \cite{desi04}  \\
25    & Psi$^1$~Draconis~A    & $-$0.10  $\pm$ 0.04  & \cite{endl16}  \\
25    & Psi$^1$~Draconis~B    & $+$0.00  $\pm$ 0.01  & \cite{endl16}  \\
\noalign{\smallskip}
\hline
\noalign{\medskip}
\multicolumn{4}{p{1.3\columnwidth}}{Note: Error bars assumed as 1$\sigma$.
$^a$Error bar estimated as $\pm$0.05.
}
\end{tabular}
\end{table*}

\begin{figure}[ht]
    \centering
	\includegraphics[width=\columnwidth]{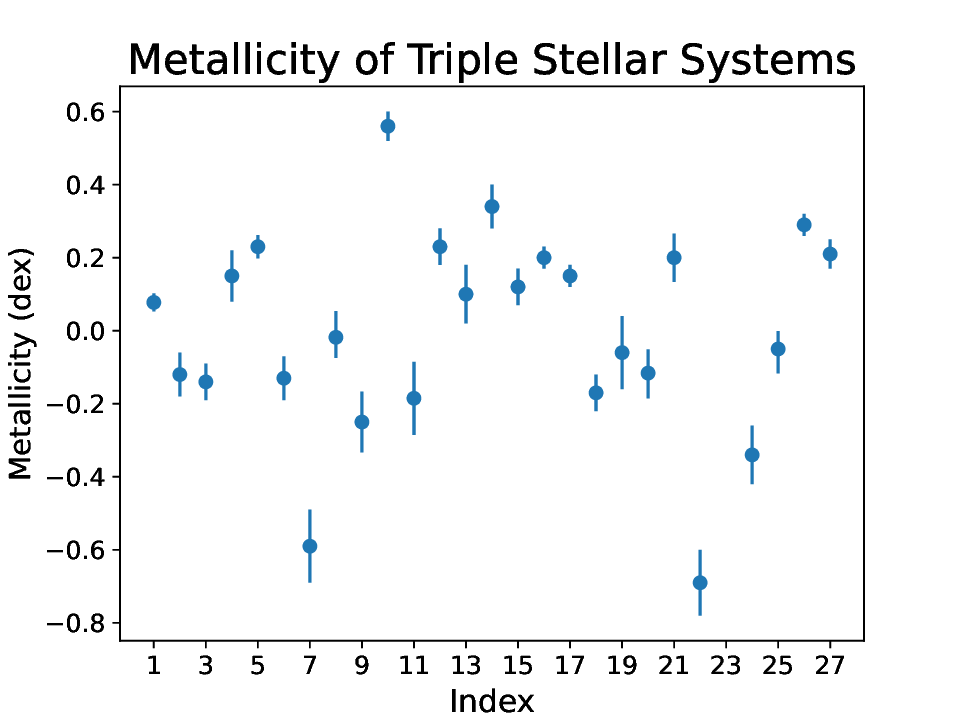}
    \caption{Metallicities of triple stellar systems, with average values
for the stellar components used if multiple determinations are available.
All error bars are 1$\sigma$.
}
\end{figure}

\begin{figure}[ht]
    \centering
	\includegraphics[width=\columnwidth]{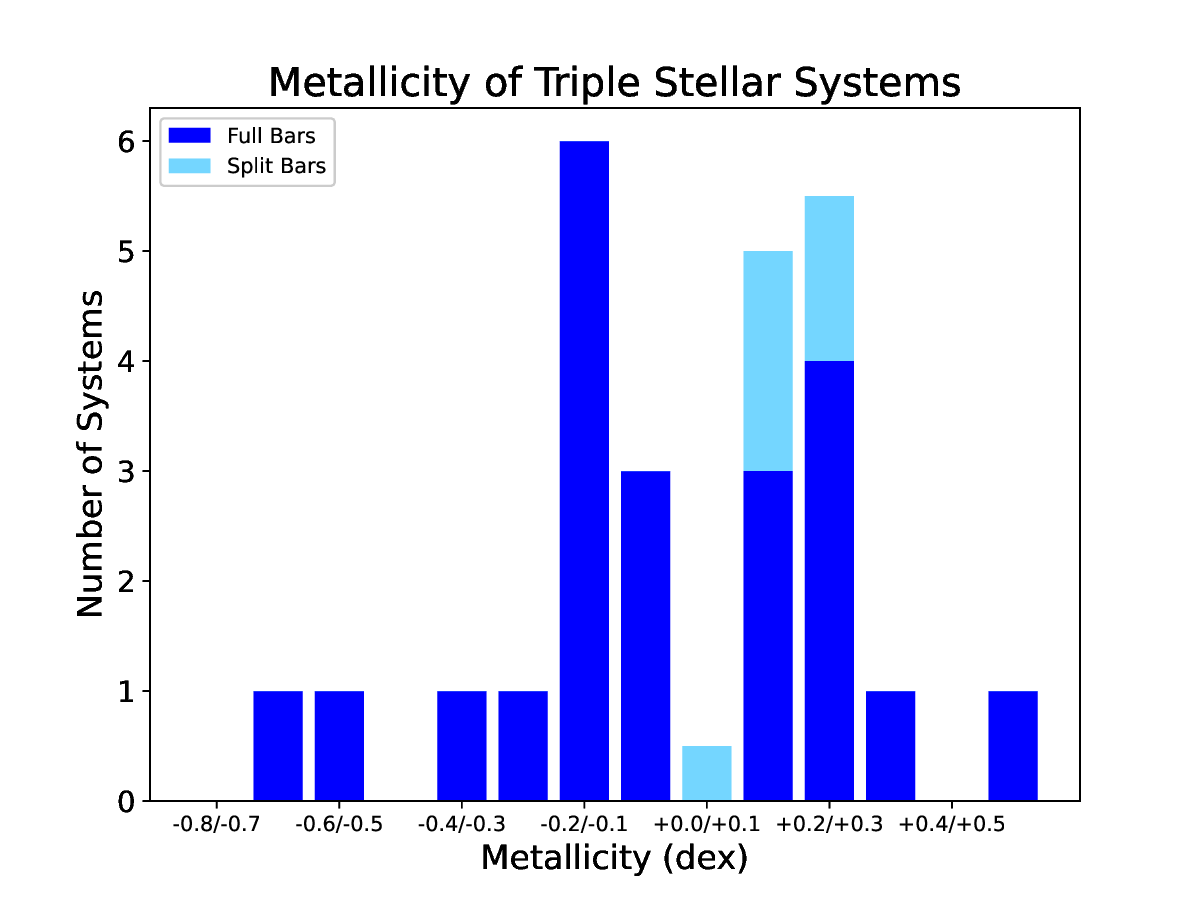}
    \caption{Histogram for the metallicity distribution of
      planet-hosting triple star systems based on the
      average values for the stellar components.  Split bars
      are used to depict border values between bins.
}
\end{figure}


\section{Statistical Analysis}

\subsection{Approach}

A pivotal aspect of this study is the computation of the statistical properties
for the age and metallicity distribution regarding the various stellar components
of the systems.  For the 27 planet-hosting triple star systems, information is
available for stars of 26 systems (see Tables 2 to 5).  The missing cases are
LTT~1445 and KOI-5 concerning the stellar age and metallicity, respectively.
Moreover, regarding the various systems, information has mostly not been obtained
for all system components.  If information is available for more than one stellar
component (which is consistently found to be very similar, as expected), average
values have been taken for the respective analyses.

The computation of the mean and median values for the distributions is highly
straightforward, noting that all 26 useable systems are given the same weight.
However, for our main analysis, we also consider the error bars for the data.
In most cases, those have been given as part of the original analyses; however,
in the small number of cases of unavailability good-faith estimates have been made.
Clearly, the consideration of the various error bars requires a detailed
numerical statistical approach; see \cite{mont06} for a description of the
methods as used.  This kind modeling assumes intrinsic Gaussian distributions
for each data point\footnote{
Note that although the error bar for the age of HD~2638 / HD~2567 \citep{bonf15}
is unusually large, it is still considered in the analysis, except that negative age 
values have been omitted (as also done in other cases if appearing as part of the
statistical simulation).}; see Table 2 and 5, including information on the respective
$1\sigma$ uncertainties.  On the other hand, the overall distributions for the
stellar ages and metallicities are noticeably non-Gaussian.

\subsection{Results}

Table 2 and 4 list the data for the stellar ages and metallicities,
respectively --- noting that in some cases information on the stellar
components other than the main component is available as well; see
Table 3 and 5.  The mean and median values for the stellar age distribution
are given as ${\cal A}_{\rm M} = 4.48$~Gyr and ${\cal A}_{\rm Mdn} = 3.26$~Gyr,
respectively.  The results for the metallicity distribution read
{\it Z}$_{\rm M} =$ $+$0.00 dex and {\it Z}$_{\rm Mdn} =$ $+$0.04 dex.

If the error bars of the data are disregarded (included here mostly
for tutorial reasons), we find similar results for both the mean and median
in terms of the metallicity distribution.  However, for the stellar age distribution
we obtain ${\cal A}_{\rm M} = 3.63$~Gyr and ${\cal A}_{\rm Mdn} = 3.45$~Gyr; these
values somewhat deviate from the previous results (see Table~6).

Both the stellar age and metallicity distributions are notably non-Gaussian.
The stellar age distribution is akin to a Poisson distribution, with an
outlier at 11.23~Gyr (Kepler-444) added on.  Moreover, the metallicity
distribution is concentrated between ${\rm [Fe/H]} = -$0.20 dex and $+$0.20 dex,
thus consistent with the solar metallicity value of ${\rm [Fe/H]} = +$0.00 dex,
even though there is a well-pronounced minimum at that data.

In order to obtain further insight into the shape and structure of those
distribution, we also have calculated the LHS and RHS deviations from the
respective medians with respect to different levels (akin to well-defined
standard deviations for Gaussian distributions); see Table~7.  These results
confirm that in general (1) planet-hosting systems are relatively young and
(2) the overall stelar metallicity values do not deviate much from the solar
case.  Section~5 provides further comments on these results, including
some interpretation while also taking into account observational findings
available in the literature.

%
\begin{table}
	\centering
	\caption{General statistics}
	\begin{tabular}{lcc}
\noalign{\smallskip}
\hline
\noalign{\smallskip}
Parameter        & Age             &  Metallicity         \\
...              & (Gyr)           &  (dex)               \\
\noalign{\smallskip}
\hline
\noalign{\smallskip}
Mean            &    4.48         &  $+$0.00             \\
Mean (NEB)      &    3.63         &  $+$0.00             \\
Median          &    3.26         &  $+$0.04             \\
Median (NEB)    &    3.45         &  $+$0.03             \\
Minimum         &    0.026        &  $-$0.69             \\
Maximum         &   11.23         &  $+$0.56             \\
\noalign{\smallskip}
\hline
\noalign{\medskip}
\multicolumn{3}{p{0.7\columnwidth}}{Note: NEB means that no error bars
have been considered for the individual data points.
}
\end{tabular}
\end{table}


%
\begin{table}
	\centering
	\caption{Distribution analysis}
	\begin{tabular}{lcc}
\noalign{\smallskip}
\hline
\noalign{\smallskip}
Parameter          & \multicolumn{2}{c}{Age (Gyr)         }  \\
\noalign{\smallskip}
\hline
\noalign{\smallskip}
Deviation                  & LHS     &  RHS     \\
\noalign{\smallskip}
\hline
\noalign{\smallskip}
50\% Level                 &  1.71   &  5.08    \\
75\% Level                 &  0.86   &  6.52    \\
95\% Level                 &  ...    & 10.59    \\
\noalign{\smallskip}
\hline
\noalign{\smallskip}
Deviation          & \multicolumn{2}{c}{Metallicity (dex)}  \\
\noalign{\smallskip}
\hline
\noalign{\smallskip}
50\% Level                 & $-$0.16  & $+$0.20  \\
75\% Level                 & $-$0.29  & $+$0.27  \\
95\% Level                 & $-$0.69  & $+$0.54  \\
\noalign{\smallskip}
\hline
\noalign{\smallskip}
\end{tabular}
\end{table}


\section{Discussion}

We pursued a detailed statistical analysis of the ages
and metallicities of stars known to be members of planet-hosting triple
star systems.  This is an exploratory follow-up study of the previous work by \cite{cunt22},
who presented a catalog of multiple-star systems known to host exoplanets.
With controversial cases omitted, 27 star-planet systems have been identified.
Observational studies indicate that all planet-hosting triple stellar systems
are highly hierarchic, consisting of a binary accompanied by a distant
stellar component, which is in orbit about the common center of mass; see
also \cite{buse18} for previous results.  Regarding the binary, both
S-type and P-type orbits are possible, whereas the distant binary
may serve as a planetary hub as well.

Regarding the data acquisition and the associated statistics, our work
followed a two-tier approach.  For the stellar ages and metallicities,
information is obtained from the literature.  Hence, the authors are
not engaged in any of those determinations but take note of existing
information for further statistical analyses, the main focus of this work.

Out of the 27 systems, age and metallicity determinations exist for
26 systems.  Error bars for both the stellar ages and metallicities are
available as well, which have been considered accordingly.  For stellar age
determinations, a variety of methods has been used; see Section 3.2 for
details.  Metallicity determinations are consistently based on quality
spectroscopy.  These values are often given as part of the respective
planet discovery article, which typically also provide detailed information
about the planetary host star.

The mean value and the median value for the stellar
age distribution are identified as 4.48~Gyr and 3.26~Gyr, respectively.
Hence, planet-hosting triple star systems are typically notably younger
than stars found in the solar neighborhood.  For example, \cite{lin18}
examined the stellar ages and masses of about 4000 stars in the
solar neighborhood based on six well-studied literature samples
using both {\it Hipparcos} and TGAS\footnote{TGAS is an acronym for
{\it Tycho-Gaia Astrometric Solution}.} parallaxes.  They report age values
of $11.18 \pm 2.57$~Gyr and $11.13 \pm 2.58$~Gyr (closely associated
with low-mass stars) in agreement with the literature values.  Similar
results have been obtained by \cite{binn00}.
In contrast, the range of stellar ages for planet-hosting triple stellar
systems, as identified in this study (see Table 6 and 7), reads
${\cal A} [\pm 50\%] = 3.26\substack{+1.82 \\ -1.55}$~Gyr and
${\cal A} [\pm 75\%] = 3.26\substack{+3.26 \\ -2.40}$~Gyr.
These numbers are significantly lower than the typical ages of stars
located in the solar neighborhood.

Though the statistical significance of this age difference might be
moderately compromised owing to the limited number of systems available,
the most plausible explanation is a double selection effect given as
(1) planet-hosting stars of those systems are relatively massive,
i.e., main-sequence stars of relatively early spectra types (F and G
rather than K and M).  As discussed by \cite{cunt22} and references
therein, a notable fraction of those stars is of spectral type
F or G, compared the spectral type K and M, while the latter make
up the vast majority of stars in the solar neighborhood and beyond
\citep[e.g.,][]{krou01,krou02,chab03}.  Consequently, planet-hosting
stars in triple stellar systems spend a reduced amount of time on the
main-sequence.
(2) Planets situated in triple stellar systems often face a limited duration
of orbital stability compared to planets hosted by binary systems or
single stars.  Many of those will be ejected eventually, thus entering
the stage of free-floating planets previously observed in our galaxy
\citep[e.g.,][]{luca00,zapa00,mire22}.

Planetary orbital stability is quite often noticeably reduced in multiple
stellar systems compared to planet-hosting single-star or binary systems
as indicated by previous analyses; see, e.g., \cite{verr07},
\cite{geor13}, \cite{corr16}, \cite{buse18}, \cite{myll18}, and
related work.  Depending on the system parameters, the
multiplicity of stellar systems may adversely impact the temporal
evolution of those systems --- with an increased level of endurance
to be expected for hierarchic systems, which may be the underlying
reasons why to date no planet-hosting non-hierarchic triple systems
have been found.

For the stellar metallicities of stars in planet-hosting triple star
systems, the median value mostly agrees with the metallicity identified
for the Sun.  At first sight, this result appears to be inconsistent
with the previous finding that planet-hosting single stars tend to be
metal-rich \citep[e.g.,][~and related work]{gonz97,gonz98,fisc05}.
For example, \cite{fisc05} carried out a spectral synthesis modeling
program for 1040 FGK-type stars.  Regarding 850 stars, they determined
that fewer than 3\% of stars with $-0.5<[$Fe/H$]<0.0$ have Doppler-detected
planets. Above solar metallicity, they identified a smooth and rapid rise
in the fraction of stars with planets. However, at [Fe/H]$>+0.3$ dex,
25\% of observed stars gas giant planets were detected.

On the other hand, the range of metallicities for stars being part of
planet-hosting triple star systems, based on this study
(see Table 6 and 7), reads
${\it Z} [\pm 50\%] = +0.04\substack{+0.16 \\ -0.20}$~dex and
${\it Z} [\pm 75\%] = +0.04\substack{+0.23 \\ -0.33}$~dex.
Thus, for those stars the width of the stellar metallicity distribution
is relatively large, thus disallowing a firm conclusion.
In addition, the shape of the distribution is highly unstructured.
Although the median value appears to indicate a solar-like metallicity,
the overall data distribution, based on small number statistics, is not
inconsistent with the previous finding that stars with planets
tend to be metal-rich.  For updated information on the physics of
star--planet metallicity connections see, e.g., \cite{wang15}, \cite{zhu19},
\cite{laug00}, \cite{osbo20}, and \cite{jian21}.


\section{Conclusions and Outlook}

In conclusion, we obtained the following results: 
(1) For the various stellar components, the age and metallicity values
agree within the error bars, as expected;
(2) on average, the stellar ages as deduced are considerably smaller than
expected from previous studies about the solar neighborhood; and (3)
the stellar metallicities are on average
solar-like; however, owing to the limited number of
data, this result is not inconsistent with the previous finding that
stars with planets tend to be metal-rich.  In fact,
the metallicity distribution as obtained has a relatively large width.

The findings of this study are relevant to both astrophysics and astrobiology.
Important aspects deserving future attention include the study of scenarios of
successful planet formation in triple star systems regarding different kinds
of stellar hierarchies, assessments of planetary orbital stability for
different types of systems, and studies of habitability, including the
general possibility of exolife.

Due to the low number of planet-hosting triple star systems known to date,
given as about one discovery per year so far, we are looking forward
to future expansions of the existing data base.  This will allow us to
revisit the currently existing statistical analysis with the intent 
of making it more robust.  It will also help us to target the various
open astrophysical and astrobiological questions associated with the
various star--planet configurations.


\section*{Acknowledgments}

This work has been supported by the Department of Physics, University of
Texas at Arlington.  In addition, the authors acknowledge assistance by
Gregory E. Luke and Lindsey Boyle about data collection, and input
by Gumseng Lamau regarding literature review.



\begin{thebibliography}{}

\bibitem[Adibekyan et al.(2012)]{adib12}
Adibekyan, V.Zh., Sousa, S.G., Santos, N.C., et al.: \aap, \ {\bf 545,} A32 (2012)

\bibitem[Anglada-Escud{\'e} et al.(2012)]{angl12}
Anglada-Escud{\'e}, G., Arriagada, P., Vogt, S.S., et al.: \apjl, \ {\bf 751,} L16 (2012)

\bibitem[Anglada-Escud{\'e} et al.(2016)]{angl16}
Anglada-Escud{\'e}, G., Amado, P.J., Barnes, J., et al.: \nat, \ {\bf 536,} 437 (2016)

\bibitem[Baines et al.(2018)]{bain18}
Baines, E.K., Armstrong, J.T., Schmitt, H.R., et al.: \aj, \ {\bf 155,} 30 (2018)

\bibitem[Baliunas et al.(1995)]{bali95}
Baliunas, S.L., Donahue, R.A., Soon, W.H., et al.:
\apj \ {\bf 438,} 269 (1995)

\bibitem[Bell et al.(2015)]{bell15}
Bell, C.P.M., Mamajek, E.E., Naylor, T.: \mnras, \ {\bf 454,} 593 (2015)

\bibitem[Bellinger et al.(2019)]{bell19}
Bellinger, E.P., Hekker, S., Angelou, G.C., Stokholm, A., Basu, S.: \aap, \ {\bf 622,} A130 (2019)

\bibitem[Binney et al.(2000)]{binn00}
Binney, J., Dehnen, W., Bertelli, G.: \mnras, \ {\bf 318}, 658 (2000)

\bibitem[Bonfanti et al.(2015)]{bonf15}
Bonfanti, A., Ortolani, S., Piotto, G., Nascimbeni, V.: \aap, \ {\bf 575,} A18 (2015)

\bibitem[Bonfanti et al.(2016)]{bonf16}
Bonfanti, A., Ortolani, S., Nascimbeni, V.: \aap, \ {\bf 585,} A5 (2016)

\bibitem[Boyajian et al.(2013)]{boya13}
Boyajian, T.S., von Braun, K., van Belle, G., et al.: \apj, \ {\bf 771,} 40 (2013)

\bibitem[Boyle \& Cuntz(2021)]{boyl21}
Boyle, L., Cuntz, M.: RNAAS, \ {\bf 5,} 285 (2021)

\bibitem[Busetti et al.(2018)]{buse18}
Busetti, F., Beust, H., Harley, C.: \aap, \ {\bf 619,} A91 (2018)

\bibitem[Campante et al.(2015)]{camp15}
Campante, T.L., Barclay, T., Swift, J.J.: \apj, \ {\bf 799,} 170 (2015)

\bibitem[Cayrel de Strobel et al.(1981)]{cayr81}
Cayrel de Strobel, G., Bentolila, C., Hauck, B., Lovy, D.: \aaps, \ {\bf 45,} 97 (1981)

\bibitem[Chabrier(2003)]{chab03}
Chabrier, G.: \pasp, \ {\bf 115,} 763 (2003)

\bibitem[Chavero et al.(2019)]{chav19}
Chavero, C., de la Reza, R., Ghezzi, L., et al.: \mnras, \ {\bf 487,} 3162 (2019)

\bibitem[Cochran et al.(1997)]{coch97}
Cochran, W.D., Hatzes, A.P., Butler, R.P., Marcy, G.W.: \apj, \ {\bf 483,} 457 (1997)

\bibitem[Correia et al.(2016)]{corr16}
Correia, A.C.M., Bou{\'e}, G., Laskar, J.: CeMDA, \ {\bf 126,} 189 (2016)

\bibitem[Cuntz(2014)]{cunt14}
Cuntz, M.: \apj, \ {\bf 780,} 14 (2014)

\bibitem[Cuntz et al.(1999)]{cunt99}
Cuntz, M., Rammacher, W., Ulmschneider, P., Musielak, Z.E., Saar, S.H.:
\apj \ {\bf 522,} 1053 (1999)

\bibitem[Cuntz et al.(2022)]{cunt22}
Cuntz, M., Luke, G.E., Millard, M.J., Boyle, L., Patel, S.D.: \apjs, \ {\bf 263,} 33 (2022)

\bibitem[Damasso et al.(2020)]{dama20}
Damasso, M., Del Sordo, F., Anglada-Escud{\'e}, G., et al.: Science Adv., \ {\bf 6,} 7467 (2020)

\bibitem[Delgado Mena et al.(2017)]{delg17}
Delgado Mena, E., Tsantaki, M., Adibekyan, V. Zh., et al.: \aap, \ {\bf 606,} A94 (2017)

\bibitem[Desidera et al.(2004)]{desi04}
Desidera, S., Gratton, R. G., Scuderi, S., et al.: \aap, \ {\bf 420,} 683 (2004)

\bibitem[Desidera et al.(2011)]{desi11}
Desidera, S., Carolo, E., Gratton, R., et al.: \aap, \ {\bf 533,} A90 (2011)

\bibitem[Domingos et al.(2015)]{domi15}
Domingos, R.C., Winter, O.C., Izidoro, A.: IJAsB, \ {\bf 14,} 153 (2015)

\bibitem[Duquennoy and Mayor(1991)]{duqu91}
Duquennoy, A., Mayor, M.: \aap \ {\bf 248,} 485 (1991)

\bibitem[Dvorak(1982)]{dvor82}
Dvorak, R.: OAWMN, \ {\bf 191,} 423 (1982)

\bibitem[Eastman et al.(2016)]{east16}
Eastman, J.D., Beatty, T.G., Siverd, R.J., et al.: \aj, \ {\bf 151,} 45 (2016)

\bibitem[Eggenberger et al.(2004)]{egge04}
Eggenberger, A., Udry, S., Mayor, M.: \aap \ {\bf 417,} 353 (2004)

\bibitem[Endl et al.(2016)]{endl16}
Endl, M., Brugamyer, E.J., Cochran, W.D., et al.: \apj, \ {\bf 818,} 34 (2016)

\bibitem[Engle(2024)]{engl24}
Engle, S.G.: \apj, \ {\bf 960,} 62 (2024)

\bibitem[Faria et al.(2022)]{fari22}
Faria, J.P., Su{\'a}rez Mascare{\~ n}o, A., Figueira, P., et al.: \aap, \ {\bf 658,} A115 (2022)

\bibitem[Feng et al.(2019)]{feng19}
Feng, F., Anglada-Escud{\'e}, G., Tuomi, M., et al.: \mnras, \ {\bf 490,} 5002 (2019)

\bibitem[Fischer \& Valenti(2005)]{fisc05}
Fischer, D.A., Valenti, J.: \apj, \ {\bf 622,} 1102 (2005)

\bibitem[Fischer et al.(2003)]{fisc03}
Fischer, D.A., Marcy, G.W., Butler, R.P., et al.: \apj, \ {\bf 586,} 1394 (2003)

\bibitem[Fuhrmann(2008)]{fuhr08}
Fuhrmann, F.: \mnras, \ {\bf 384,} 173 (2008)

\bibitem[Georgakarakos(2013)]{geor13}
Georgakarakos, N.: New Astronomy, \ {\bf 23-24,} 41 (2013)

\bibitem[Gonzalez(1997)]{gonz97}
Gonzalez, G.: \mnras, \ {\bf 285}, 403 (1997)

\bibitem[Gonzalez(1998)]{gonz98}
Gonzalez, G.: \aap, \ {\bf 334}, 221 (1998)

\bibitem[Guillout et al.(2009)]{guil09}
Guillout, P., Klutsch, A., Frasca, A., et al.: \aap, \ {\bf 504,} 829 (2009)

\bibitem[Hartman et al.(2015)]{hart15}
Hartman, J.D., Bakos, G.{\'A}., Buchhave, L.A.: \aj, \ {\bf 150,} 197 (2015)

\bibitem[Hauser \& Marcy(1999)]{haus99}
Hauser, H.M., Marcy, G.W.: \pasp, \ {\bf 111,} 321 (1999)

\bibitem[Hebb et al.(2009)]{hebb09}
Hebb, L., Collier Cameron, A., Loeillet, B., et al.: \apj, \ {\bf 693,} 1920 (2009)

\bibitem[Hirsch et al.(2017)]{hirs17}
Hirsch, L.A., Ciardi, D.R., Howard, A.W., et al.: \aj, \ {\bf 153,} 117 (2017)

\bibitem[Hjorth et al.(2019)]{hjor19}
Hjorth, M., Justesen, A.B., Hirano, T., et al.: \mnras, \ {\bf 484,} 3522 (2019)

\bibitem[Howard et al.(2010)]{howa10}
Howard, A.W., Johnson, J.A., Marcy, G.W., et al.: \apj, \ {\bf 721,} 1467 (2010)

\bibitem[Jiang et al.(2021)]{jian21}
Jiang, J., Zhao, D., Ji, X., Xie, B., Fahy, K.A.: Universe, \ {\bf 7,} 88 (2021)

\bibitem[Jofr{\'e}l et al.(2015)]{jofr15}
Jofr{\'e}l, E., Petrucci, R., Saffe, C., et al.: \aap, \ {\bf 574,} A50 (2015)

\bibitem[Jones et al.(2002)]{jone02}
Jones, H.R.A., Butler, R.P., Marcy, G.W., et al.: \mnras, \ {\bf 337,} 1170 (2002)

\bibitem[Joyce \& Chaboyer(2018)]{joyc18}
Joyce, M., Chaboyer, B.: \apj, \ {\bf 864,} 99 (2018)

\bibitem[Kervella et al.(2017)]{kerv17}
Kervella, P., Bigot, L., Gallene, A., Th{\'e}venin, F.: \aap, \ {\bf 597,} A137 (2017)

\bibitem[Knutson et al.(2014)]{knut14}
Knutson, H.A., Fulton, B.J., Montet, B.T., et al.: \apj, \ {\bf 785,} 126 (2014)

\bibitem[Kroupa(2001)]{krou01}
Kroupa, P.: \mnras, \ {\bf 322,} 231 (2001)

\bibitem[Kroupa(2002)]{krou02}
Kroupa, P.: Science, \ {\bf 295,} 82 (2002)

\bibitem[Latham et al.(2009)]{lath09}
Latham, D.W., Bakos, G.{\'A}., Torres, G., et al.: \apj, \ {\bf 704,} 1107 (2009)

\bibitem[Laughlin(2000)]{laug00}
Laughlin, G.: \apjl, \ {\bf 545}, L1064 (2000)

\bibitem[Lin et al.(2018)]{lin18}
Lin, J., Dotter, A., Ting, Y.-S., Asplund, M.: \mnras, \ {\bf 477}, 2966 (2018)

\bibitem[Lucas \& Roche(2000)]{luca00}
Lucas, P.W., Roche, P.F.: \mnras, \ {\bf 314}, 858 (2000)

\bibitem[Luck(2017)]{luck17}
Luck, R.E.: \aj, \ {\bf 153,} 21 (2017)

\bibitem[Mamajek \& Hillenbrand(2008)]{mama08}
Mamajek, E.E., Hillenbrand, L.A.: \apj, \ {\bf 687,} 1264 (2008)

\bibitem[Mancini et al.(2013)]{manc13}
Mancini, L., Southworth, J., Ciceri, S., et al.: \aap, \ {\bf 551,} A11 (2013)

\bibitem[Massarotti et al.(2008)]{mass08}
Massarotti, A., Latham, D.W., Stefanik, R.P., Fogel, J.: \aj, \ {\bf 135,} 209 (2008)

\bibitem[Mayor et al.(2004)]{mayo04}
Mayor, M., Udry, S., Naef, D., et al.: \aap, \ {\bf 415,} 391 (2004)

\bibitem[Macintosh et al.(2015)]{maci15}
Macintosh, B., Graham, J.R., Barman, T., et al.: Science, \ {\bf 350,} 64 (2015)

\bibitem[Metcalfe et al.(2015)]{metc15}
Metcalfe, T.S., Creevey, O.L., Davies, G.R.: \apjl, \ {\bf 811,} L37 (2015)

\bibitem[Miret-Roig et al.(2022)]{mire22}
Miret-Roig, N., Bouy, H., Raymond, S.N., et al.: \nat Astron., \ {\bf 6,} 89 (2022)

\bibitem[Mitchell et al.(2003)]{mitc03}
Mitchell, D.S., Frink, S., Quirrenbach, A., et al.: \baas, \ {\bf 35,} 1234 (2003)

\bibitem[Montgomery et al.(2006)]{mont06}
Montgomery, D.C., Peck, E.A., Vining, G.G.: Introduction to Linear Regression Analysis,
4th edn., Wiley Interscience, Hoboken, NJ (2006)

\bibitem[Morel(2018)]{more18}
Morel, T.: \aap, \ {\bf 615,} A172 (2018)

\bibitem[Mortier et al.(2013)]{mort13}
Mortier, A., Santos, N.C., Sousa, S.G., et al.: \aap, \ {\bf 558,} A106 (2013)

\bibitem[Moutou et al.(2005)]{mout05}
Moutou, C., Mayor, M., Bouchy, F., et al.: \aap, \ {\bf 439,} 367 (2005)

\bibitem[Moutou et al.(2006)]{mout06}
Moutou, C., Loeillet, B., Bouchy, F., et al.: \aap, \ {\bf 458,} 327 (2006)

\bibitem[Mugrauer et al.(2007)]{mugr07}
Mugrauer, M., Seifahrt, A., Neuh\"auser, R.: \mnras, \ {\bf 378,} 1328 (2007)

\bibitem[Myll\"ari et al.(2018)]{myll18}
Myll\"ari, A., Valtonen, M., Pasechnik, A., Mikkola, S.: \mnras, \ {\bf 476,} 830 (2018)

\bibitem[Narang et al.(2021)]{nara21}
Narang, M., Manoj, P., Ishwara Chandra, C.H., et al.: \mnras, \ {\bf 500,} 4818 (2021)

\bibitem[Nielsen et al.(2016)]{niel16}
Nielsen, E.L., De Rosa, R.J., Wang, J., et al.: \aj, \ {\bf 152,} 175 (2016)

\bibitem[Noyes et al.(1984)]{noye84}
Noyes, R.W., Hartmann, L.W., Baliunas, S.L., Duncan, D.K., Vaughan, A.H.:
\apj \ {\bf 279,} 763 (1984)

\bibitem[Osborn et al.(2020)]{osbo20}
Osborn, A., Bayliss, D.: \mnras, \ {\bf 491,} 4481 (2020)

\bibitem[Pace et al.(2003)]{pace03}
Pace, G., Pasquini, L., Ortolani, S.: \aap, \ {\bf 401,} 997 (2003)

\bibitem[Patience et al.(2002)]{pati02}
Patience, J., White, R.J., Ghez, A.M., et al.: \apj \ {\bf 581,} 654 (2002)

\bibitem[Raghavan et al.(2006)]{ragh06}
Raghavan, D., Henry, T.J., Mason, B.D., et al.: \apj \ {\bf 646,} 523 (2006)

\bibitem[Raghavan et al.(2010)]{ragh10}
Raghavan, D., McAlister, H.A., Henry, T.J., et al.: \apjs \ {\bf 190,} 1 (2010)

\bibitem[Rajan et al.(2017)]{raja17}
Rajan, A., Rameau, J., De Rosa, R.J., et al.: \aj, \ {\bf 154,} 10 (2017)

\bibitem[Roell et al.(2012)]{roel12}
Roell, T., Neuh\"auser, R., Seifahrt, A., Mugrauer, M.: \aap \ {\bf 542,} A92 (2012)

\bibitem[Saffe et al.(2005)]{saff05}
Saffe, C., G{\'o}mez, M., Chavero, C.: \aap, \ {\bf 443,} 609 (2005)

\bibitem[Schlaufman \& Laughlin(2010)]{schl10}
Schlaufman, K.C., Laughlin, G.: \aap, \ {\bf 519,} A105 (2010)

\bibitem[Shporer et al.(2011)]{shpo11}
Shporer, A., Jenkins, J.M., Rowe, J.F., et al.: \aj, \ {\bf 142,} 195 (2011)

\bibitem[Shporer et al.(2014)]{shpo14}
Shporer, A., O'Rourke, J.G., Knutson, H.A., et al.: \apj, \ {\bf 788,} 92 (2014)

\bibitem[Sloane et al.(2023)]{sloa23}
Sloane, S.A., Guinan, E.F., Engle, S.G.: RNAAS, \ {\bf 7,} 135 (2023)

\bibitem[Sousa et al.(2018)]{sous18}
Sousa, S.G., Adibekyan, V., Delgado-Mena, E., et al.: \aap, \ {\bf 620,} 58 (2018)

\bibitem[Southworth et al.(2020)]{sout20}
Southworth, J., Bohn, A.J., Kenworthy, M.A., Ginski, C., Mancini, L.: \aap, \ {\bf 635,} A74 (2020)

\bibitem[Th{\'e}venin et al.(2002)]{thev02}
Th{\'e}venin, F., Provost, J., Morel, P., et al.: \aap, \ {\bf 392,} L9 (2002)

\bibitem[Tsantaki et al.(2013)]{tsan13}
Tsantaki, M., Sousa, S.G., Adibekyan, V.Zh., et al.: \aap, \ {\bf 555,} A150 (2013)

\bibitem[Tucci Maia et al.(2014)]{tucc14}
Tucci Maia, M., Mel{\'e}ndez, J., Ram{\'{\i}}rez, I.: \apjl, \ {\bf 790,} L25 (2014)

\bibitem[Queloz et al.(2010)]{quel10}
Queloz, D., Anderson, D.R., Collier Cameron, A., et al.: \aap, \ {\bf 517,} L1 (2010)

\bibitem[Valenti \& Fischer(2005)]{vale05}
Valenti, J.A., Fischer, D.A.: \apjs, \ {\bf 159,} 141 (2005)

\bibitem[Verrier \& Evans(2007)]{verr07}
Verrier, P.E., Evans, N.W.: \mnras, \ {\bf 382,} 1432 (2007)

\bibitem[Viani et al.(2018)]{vian18}
Viani, L.S., Basu, S., Joel Ong, J.M., Bonaca, A., Chaplin, W.J.:
\apj, \ {\bf 858,} 28 (2018)

\bibitem[Wang \& Fischer(2015)]{wang15}
Wang, J., Fischer, D.A.: \aj, \ {\bf 149}, 14 (2015)

\bibitem[Wang et al.(2021)]{wang21}
Wang, X.-Y., Wang, Y.-H., Wang, S., et al.: \apjs, \ {\bf 255,} 15 (2021)

\bibitem[Winters et al.(2019)]{wint19}
Winters, J.G., Medina, A.A., Irwin, J.M., et al.: \aj, \ {\bf 158,} 152 (2019)

\bibitem[Zapatero Osorio et al.(2000)]{zapa00}
Zapatero Osorio, M.R., B{\'e}jar, V.J.S. ; Mart{\'{\i}}n, E.L., et al.: Science, \ {\bf 290,} 103 (2000)

\bibitem[Zhu(2019)]{zhu19}
Zhu, W.: \apj, \ {\bf 873}, 8 (2019)

\bibitem[Zucker et al.(2002)]{zuck02}
Zucker, S., Naef, D., Latham, D.W., et al.: \apj, \ {\bf 568,} 363 (2002)

\bibitem[Zucker et al.(2004)]{zuck04}
Zucker, S., Mazeh, T., Santos, N.C., Udry, S., Mayor, M.: \aap, \ {\bf 426,} 695 (2004)

\end{thebibliography}
\end{document}